# Large Amplitude Switchback Turbulence: Possible Magnetic Velocity Alignment Structures

Honghong Wu[1], Chuanyi Tu[1], Xin Wang[2], and Liping Yang[3]
[1] School of Earth and Space Sciences, Peking University, Beijing, People's Republic of China; honghongwu@pku.edu.cn
[2] School of Space and Environment, Beihang University, Beijing, People's Republic of China
[3] SIGMA Weather Group, State Key Laboratory for Space Weather, National Space Science Center, Chinese Academy of Sciences, Beijing, People's Republic of China

## Abstract

Switchbacks are widely acknowledged phenomena observed by the Parker Solar Probe and appear to occur in patches. Previous studies focused on the fluctuations at the magnetic reversals. However, the nature of the fluctuations inside the switchbacks remains unknown. Here we utilize the magnetic field data and plasma data measured by the Parker Solar Probe in the first four encounters. We investigate the fluctuations in the switchback intervals of 100 s with $B_R > 0$ at every instant and compare them to the fluctuations in the nonswitchback intervals of 100 s with $\theta_{RB} > 160°$ at every instant. We calculate normalized cross-helicity $\sigma_c$, normalized residual energy $\sigma_r$, correlation coefficient $C_{vb}$ between $\delta \mathbf{v}_A$ and $\delta \mathbf{v}$, Alfvén ratio $r_A$, and the amplitude of magnetic and kinetic fluctuations. We find that the switchback intervals exhibit a distribution of $\sigma_c$ similar with the nonswitchback intervals. However, the $r_A$ of switchback intervals is around 0.35, while the nonswitchback intervals have $r_A$ around 0.65, indicating the fluctuations in the switchbacks are more magnetically dominated. We also find that the distribution pattern of pixel average amplitude of both $\delta \mathbf{v}_A$ and $\delta \mathbf{v}$ of switchback intervals in the $C_{vb}$–$\sigma_r$ plane show a vertical stripe feature at $C_{vb} > 0.8$, illustrating the possible magnetically dominant magnetic-velocity alignment structure. These results will help us to understand the nature and the formation of the switchback turbulence.

*Unified Astronomy Thesaurus concepts:* Solar wind (1534); Interplanetary turbulence (830)

## 1. Introduction

Ubiquitous magnetic field deflections described as "switchbacks" were observed by the Parker Solar Probe (PSP) during its encounter phases (Bale et al. 2019; Kasper et al. 2019). The switchbacks sometimes show a full reversal of the radial component of the magnetic field and then return to "regular" conditions (Borovsky 2020; Krasnoselskikh et al. 2020). Mozer et al. (2020) showed enhanced wave power at ~0.3–10 Hz at the boundaries and inside switchbacks and proposed that these waves may result from large velocity shears at the switchback boundaries. Krasnoselskikh et al. (2020) found that switchbacks have a narrow boundary layer that accommodates flowing currents and Farrell et al. (2020) demonstrated that the boundary transition into and out of the switchbacks was abrupt.

Switchbacks appear to occur in patches on the scale of hours to days, separated by intervals of "quieter" radial fields (Horbury et al. 2020). Mozer et al. (2020) analyzed switchback evolution, content, and plasma effects and found that the switchbacks contain enhanced solar wind bulk flow, Poynting flux, and thermal energy. Woolley et al. (2020) found that the proton core parallel temperature was very similar inside and outside of switchbacks using Solar Probe Cup measurements and concluded that switchbacks were consistent with Alfvénic pulses traveling along open magnetic field lines. Using data measured by the SPAN-Ai instrument, Woodham et al. (2020) observed that the patched switchbacks were correlated with enhancements in parallel temperature and they interpreted the patches were embedded within a large-scale structure, which was consistent with formation by reconnection-associated mechanisms in the corona. Alfvénic switchbacks observed by PSP are suggested to evolve from Alfvén pulses as generated and launched from interchange reconnection with discontinuous guide field (He et al. 2020).

McManus et al. (2020) identified that the fluctuations had an inward propagating sense during the field reversal. Chaston et al. (2020) reported the dominance of a broad spectrum propagating antiparallel to the background magnetic field while the magnetic reversals from inward to outward radial orientations provided outward and inward propagation in the plasma frame. The probability density function (PDF) of propagation directions in the wavenumber space ($k_\parallel$, $k_\perp$) is found to be similar inside and outside switchback intervals: both types of intervals show a transition of anti-quasi-parallel propagation at large scales to quasi-perpendicular propagation at small scales (Zhu et al. 2020). The above works support the idea that switchbacks are localized twists in the magnetic field and not polarity reversals. Wu et al. (2017) identified localized twists based on the combination of the propagating direction of the fluctuations and the pitch angle of strahl electrons using WIND measurements at 1 au, suggesting that local twist switchbacks exist up to 1 au. The nature and the formation of the switchbacks are important to understand the evolution of solar wind in the inner heliosphere (Macneil et al. 2020).

Dudok de Wit et al. (2020) found that the distributions of magnetic deflections were continuous and monotonically decreasing and these deflections tended to aggregate and possess "long memory," indicating that a switchback is statistically distinct from turbulence in the quiescent solar wind (Zank et al. 2020). The turbulent state is crucial to understanding the nature of switchbacks. Bourouaine et al.

(2020) performed a study on turbulence characteristics of switchback and nonswitchback intervals between 2018 November 3 and 2018 November 9 based on the conditioned correlation functions technique. They selected the switchbacks with outward radial magnetic field and collected these data points as a discontinuous interval. They obtained the power spectra from this discontinuous interval. Based on the spectra, they show the turbulent characteristics including spectra of normalized cross-helicity ($\sigma_c$) and normalized residual energy ($\sigma_r$) for switchbacks, and found that $\sigma_c$ is positive but smaller than $\sigma_c$ of the nonswitchbacks, and $\sigma_r$ is negative. They concluded that the switchbacks are more balanced in terms of energy for Elsässer variables.

In our study, we perform a statistical analysis on these turbulent state parameters, including not only normalized cross-helicity $\sigma_c$, normalized residual energy $\sigma_r$, but also correlation coefficient $C_{vb}$ between $\delta v_A$ and $\delta v$, Alfvén ratio $r_A$, and the amplitude of magnetic and velocity fluctuations of the switchbacks and compare them with those of nonswitchbacks. We select switchback intervals with a duration of 100 s and calculate the above parameters separately. The distribution of $\sigma_c$ of switchbacks shows no difference from that of nonswitchbacks, reaching a different conclusion from Bourouaine et al. (2020), in which one discontinuous interval was analyzed. We also find the switchbacks have aligned magnetic and kinetic fluctuations and the magnetic fluctuations are dominant. We illustrate the vertical stripe feature in the distribution pattern of pixel average amplitude of both $\delta v_A$ and $\delta v$ of switchback intervals in the $C_{vb}$–$\sigma_r$ plane. Based on those results, we propose that switchbacks may be the magnetic-velocity alignment structures with low Alfvén ratio. This paper is organized as follows. In Section 2, we describe PSP measurements and the criteria to select the switchbacks and nonswitchbacks. We introduce these parameters of turbulence state. In Section 3, we show the distributions of these parameters and analyze their implications. In Section 4, we discuss our results and draw our conclusions.

## 2. Data and Method

The PSP mission operates at the highest sampling rates in the encounter phase where the spacecraft is closer than 0.25 au (Fox et al. 2016). Here we use measurements from the encounter phase of the first four orbits of PSP. The fluxgate magnetometer (MAG) in the FIELDS instrument suite (Bale et al. 2016) provides the magnetic field data $\boldsymbol{B}$ and the Solar Probe Cup (SPC; Case et al. 2020) in the Solar Wind Electrons, Protons, and Alphas (SWEAP; Kasper et al. 2016) instrument suite provides the plasma data. Proton velocity distribution function (VDF) moments are used, including proton density $n_p$ and solar wind velocity $\boldsymbol{V}$. In this study we fit plasma and magnetic field data to a uniform time grid with the resolution of 0.8738 s.

The left panel of Figure 1 shows an interval of 100 s on 2018 November 5. During this interval, the radial component of the magnetic field $B_R > 0$ at every instant. Figure 1 (a1)–(c1) shows the magnetic field fluctuations $\delta v_A$ (blue) in Alfvén units and the velocity fluctuations $\delta v$ (red) in RTN coordinates (where $R$ is the direction from the Sun to the spacecraft, $T$ is the cross product of the solar rotation axis, and $R$, $N$ completes the right-handed coordinates.) Figure 1(d1) shows the proton number density. $\delta v_A$ and $\delta v$ are obtained from the time series subtracted by their time averages as

$$\delta v_A(t) = V_A(t) - V_{A0}, \quad (1)$$
$$\delta v(t) = V(t) - V_0, \quad (2)$$

$V_A$ is the magnetic field in Alfvén units. $V_A = B/\sqrt{4\pi n_{p0} m_p}$, where $n_{p0}$ is the time average of the proton number density and $m_p$ is the proton mass. Figure 1 (e1) shows both the Alfvén speed $V_A$ and the radial component of velocity $V_R$. $V_R$ is about three times larger than $V_A$, which is reasonable since PSP was outside the Alfvén critical point (where $V_A = V_R$) during the first four encounters. Figure 1 (f1) displays the angle $\theta_{RB}$ between the radial direction and the magnetic field. $\theta_{RB}$ is around 90°, indicating the magnetic fields are nearly perpendicular to the direction of solar wind flows. Since the quiet solar wind flows with prominently radial magnetic field (Bale et al. 2019) and the dramatic rapid polarity reversals are hallmark of switchbacks (Dudok de Wit et al. 2020), the requirement of $B_R > 0$ at every instant ensures that the interval is inside the switchback. This kind of interval belongs to switchback intervals. In the right panels of Figure 1, we also show an interval of 100 s with $\theta_{RB} > 160°$ at every instant on 2018 November 5. In this interval, the magnetic fields are nearly parallel to the direction of solar wind flows. This kind of interval belongs to nonswitchback intervals.

The normalized cross-helicity $\sigma_c$, normalized residual energy $\sigma_r$, correlation coefficient $C_{vb}$ between $\delta v_A$ and $\delta v$, and Alfvén ratio $r_A$ are all dimensionless parameters to describe the state of solar wind turbulence (Tu & Marsch 1995; Wang et al. 2020). They are defined as

$$\sigma_c = \frac{2\langle \delta v \cdot \delta v_A \rangle}{\langle \delta v^2 \rangle + \langle \delta v_A^2 \rangle}, \quad (3)$$

$$\sigma_r = \frac{\langle \delta v^2 \rangle - \langle \delta v_A^2 \rangle}{\langle \delta v^2 \rangle + \langle \delta v_A^2 \rangle}, \quad (4)$$

$$C_{vb} = \frac{\langle \delta v \cdot \delta v_A \rangle}{\sqrt{\langle \delta v^2 \rangle \langle \delta v_A^2 \rangle}}, \quad (5)$$

$$r_A = \frac{\langle \delta v^2 \rangle}{\langle \delta v_A^2 \rangle}. \quad (6)$$

Here the angled-bracket $\langle \rangle$ denotes an ensemble time average. $\sigma_r$ is related to $r_A$ by $\sigma_r = (r_A - 1)/(r_A + 1)$, and one can obtain $\sigma_c$ from $\sigma_c = C_{vb}\sqrt{1 - \sigma_r^2}$, and $\sigma_c^2 + \sigma_r^2 \leqslant 1$. $C_{vb}$ and $r_A$ are independent of each other while $\sigma_c$ and $\sigma_r$ are not (Wang et al. 2020). The values of $C_{vb}$ and $r_A$ are shown in the title of Figures 1 (a1) and (a2). We can see that for the two shown intervals, both have a relatively high correlation between the magnetic field fluctuations $\delta v_A$ and the velocity fluctuations $\delta v$ with $C_{vb} = 0.75$ and 0.77, respectively. However, the left switchback interval shows that $\delta v_A$ is dominant over $\delta v$ with $r_A = 0.39$ while the right nonswitchback interval shows a relative equivalence between $\delta v_A$ and $\delta v$ with $r_A = 1.01$. It can be seen that the fluctuations are stronger in the switchback interval than those in the nonswitchback interval, which are mainly in the perpendicular direction with respect to the magnetic field.

Figure 2 shows time series of radial magnetic field during four encounters of PSP in four panels. We can clearly see that during the four encounters, the magnetic field is essentially with clear sunward polarity, except after the heliospheric current sheet on 2020 February 1 (Zhao et al. 2020). This regularity is continuously interrupted by switchbacks (Dudok de Wit et al. 2020). There are a large number of dramatic rapid polarity reversals in the radial $B_R$ and these are hallmark of switchbacks (Dudok de Wit et al. 2020). In order to investigate the properties in the switchbacks, we select all those intervals lasting for 100 s with $B_R > 0$ at every instant during the first four encounter phases without overlapping. Dudok de Wit et al. (2020) found that the distribution of magnetic field deflections is continuous and there does not exist a typical range of deflection angles. Farrell et al. (2020) found that there are sharp switchback edges at both the entry and exit of the switchback. Using $B_R > 0$ as a criterion to select the switchbacks, on one hand, avoids the ambiguous and unclear deflections, and on the other hand, excludes the influence of the switchback boundaries, which show abrupt transitions. We set the criterion of $\delta n_p / n_{p0} < 0.1$. Here $\delta n_p$ is the rms of $n_p$. Less than 1% switchback intervals are excluded due to this criterion.

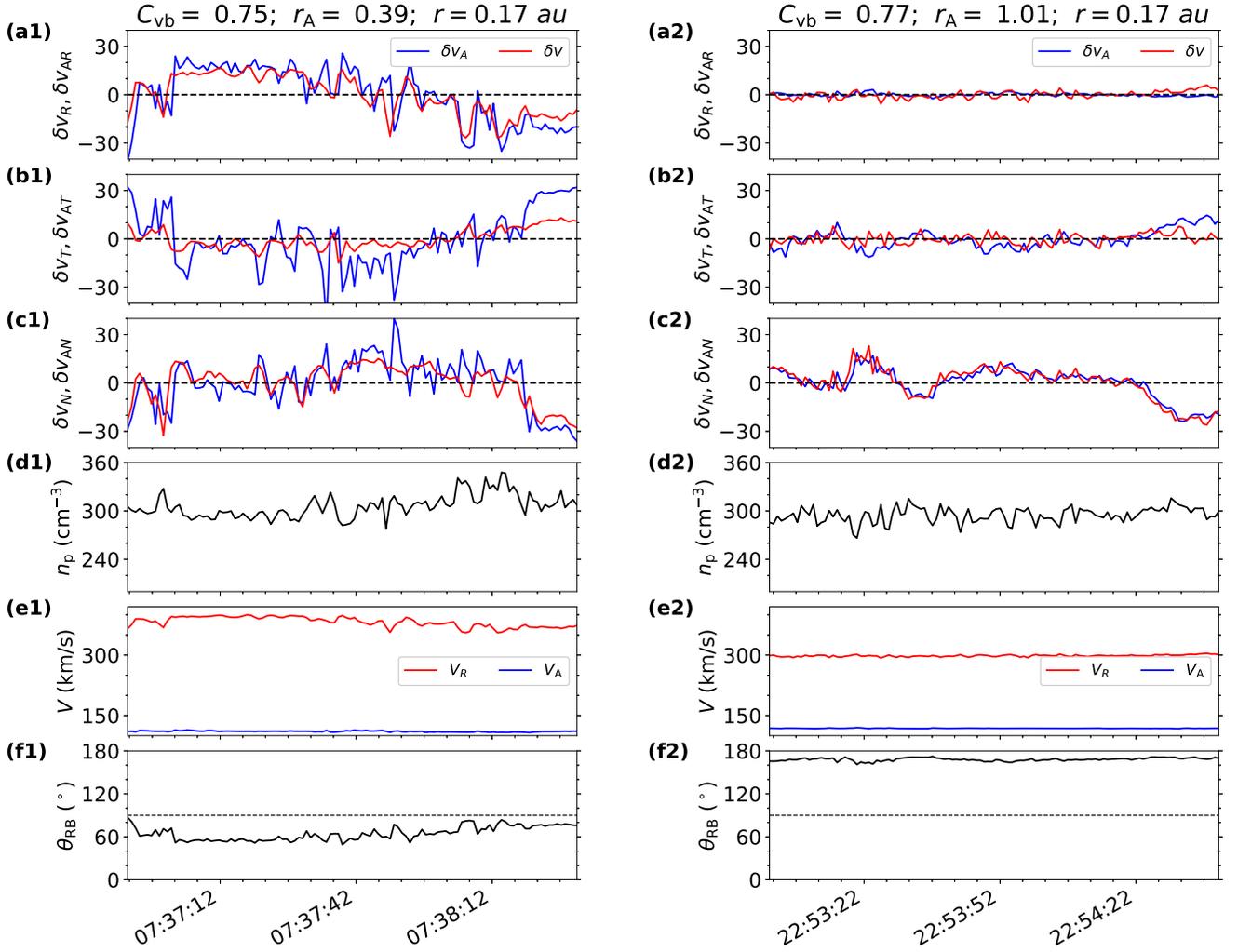

**Figure 1.** Left: a 100 s interval with $B_R > 0$ at every instant on 2018 November 5 (the switchback interval). Panels from top to bottom show the time series with a time resolution of 0.8738 s of (a1–c1) magnetic field fluctuations $\delta v_A$ (blue) in Alfvén units and velocity fluctuations $\delta v$ (red) in RTN coordinates, (d1) proton number density $n_p$, (e1) proton radial speed $V_R$ (red) and Alfvén speed $V_A$ (blue), and (f1) the angle $\theta_{RB}$ between the radial direction and the magnetic field. The correlation coefficient $C_{vb}$, Alfvén ratio $r_A$, and the radial distance $r$ are given in the title. Right: a 100 s interval with $\theta_{RB} > 160°$ at every instant on 2018 November 5 (the nonswitchback interval), with the same format as the left panels.

After the constraint of a 10% data gap, we obtain 303 switchbacks for further studies, which are denoted in red in

Figure 2. For comparison, we also analyze 2215 quiet solar wind intervals of 100 s with $\theta_{RB} > 160°$ and $\delta n_p / n_{p0} < 0.1$ at every instant. About 15% nonswitchback intervals are excluded based on the criterion $\delta n_p / n_{p0} < 0.1$. The selected nonswitchbacks are shown in blue. Note that there are no selected intervals from 2019 August 31 to 2019 September 8 and from 2020 January 29 to 2020 January 31, since there is no plasma data from SPC in those periods. We do not select intervals after the heliospheric current sheet on 2020 February 1 (Zhao et al. 2020), which leads to an antisunward polarity afterwards and thus $B_R > 0$ does not ensure that the interval is inside the switchbacks any more.

Figure 3 presents the probability distribution of $\theta_{RB}$ for both the switchback intervals and the nonswitchback intervals. $\theta_{RB}$ of the selected switchback intervals spreads from 40° to 90° with a mean value of 61°.4, suggesting a nearly perpendicular sample detection, while $\theta_{RB}$ of the selected nonswitchback intervals indicates a nearly parallel sample detection.

For the 303 selected switchback intervals and 2215 nonswitchback intervals, we calculate normalized cross-helicity $\sigma_c$, normalized residual energy $\sigma_r$, correlation coefficient $C_{vb}$ between $\delta v_A$ and $\delta v$, and Alfvén ratio $r_A$ for each interval.

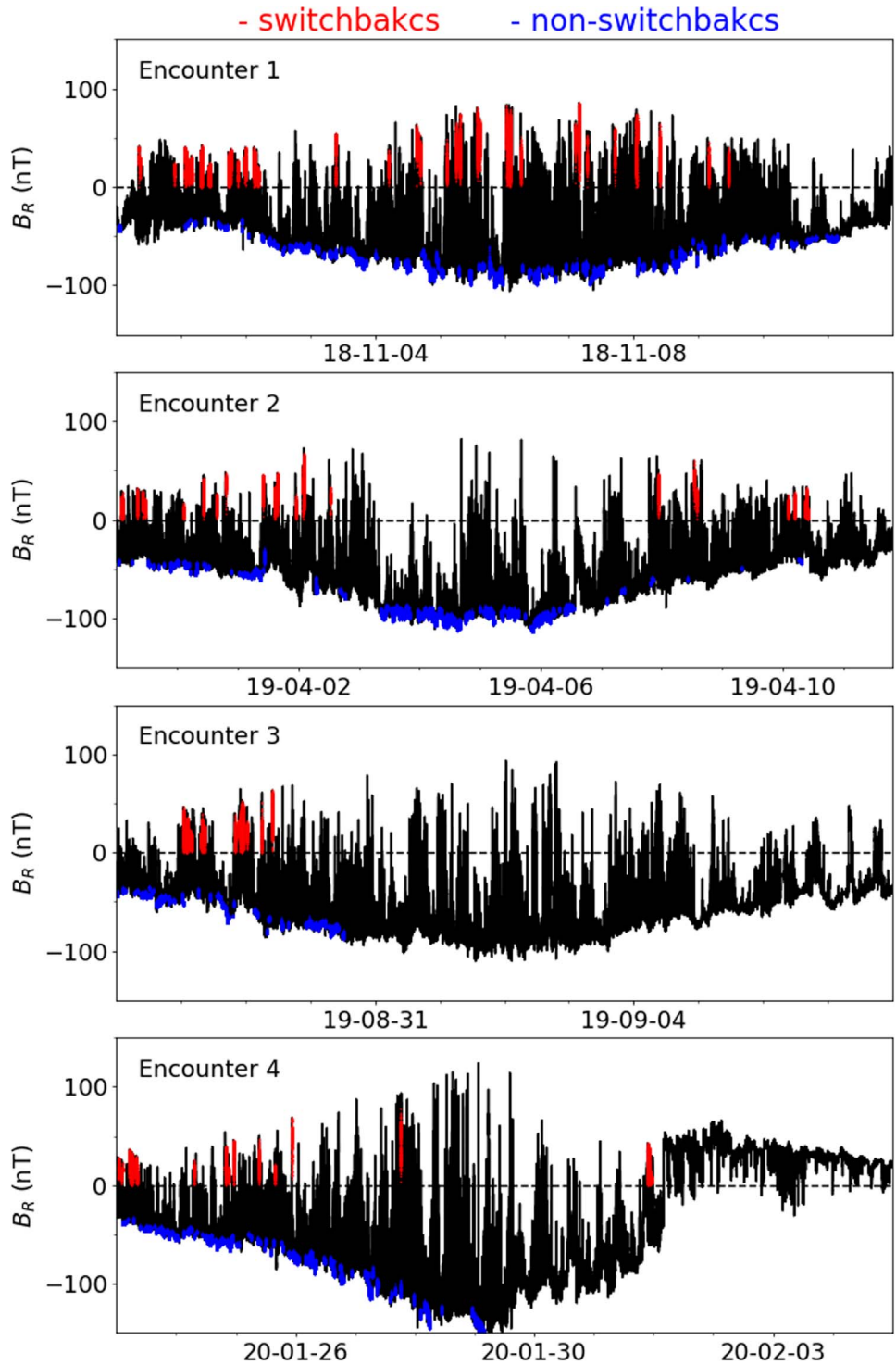

**Figure 2.** Time series of the radial magnetic field component (black) observed by PSP during the first four encounters. The 303 selected switchback intervals with $B_R > 0$ at every instant and lasting for 100 s are shown in red. The 2215 selected nonswitchback intervals with $\theta_{RB} > 160°$ at every instant and lasting for 100 s are shown in blue. There are no intervals from 2019 August 31 to 2019 September 8 and from 2020 January 29 to 2020 January 31, since there is a lack of plasma data from SPC. We do not select intervals after the heliospheric current sheet on 2020 February 1.

Furthermore, we calculate the amplitudes of both $\delta v_A$ and $\delta v$ from the square root of the variance matrix trace and obtain their pixel average and pixel standard error in the $C_{vb}$–$\sigma_r$ plane. The results are presented in the next section.

### 3. Results

Figure 4 shows the number histograms in the ($\sigma_c$, $\sigma_r$) plane for the switchback intervals in the left panel and for the nonswitchback intervals in the right panel. Almost all switchback intervals (301 out of 303) have $\sigma_c > 0$, which means that the propagating direction is antiparallel to the background magnetic field. Combining the criterion $B_R > 0$, which means that the background magnetic field is pointing outward, we conclude that the fluctuations in the switchbacks are inward propagating. Most all nonswitchbacks (98.6%) have $\sigma_c > 0$, indicating the fluctuations in 98.6% nonswitchbacks are outward propagating since they have $B_R < 0$. The inward propagation direction for switchbacks and outward propagation direction for nonswitchbacks in the plasma frame, support the idea that switchbacks are localized twists in the magnetic field (McManus et al. 2020; Chaston et al. 2020). The peak of switchback distribution centers around $\sigma_c \sim 0.7$ and $\sigma_r \sim -0.45$, indicating the magnetic field fluctuations are dominant over the velocity fluctuations. For comparison, the peak of nonswitchback distribution centers around $\sigma_c \sim 0.7$ and $\sigma_r \sim 0.1$, suggesting their Alfvénic nature. Distinct distributions in the $\sigma_c$–$\sigma_r$ plane strongly suggest that the switchback turbulence has a different Alfvénic nature with respect to the Alfvén ratio from the nonswitchbacks.

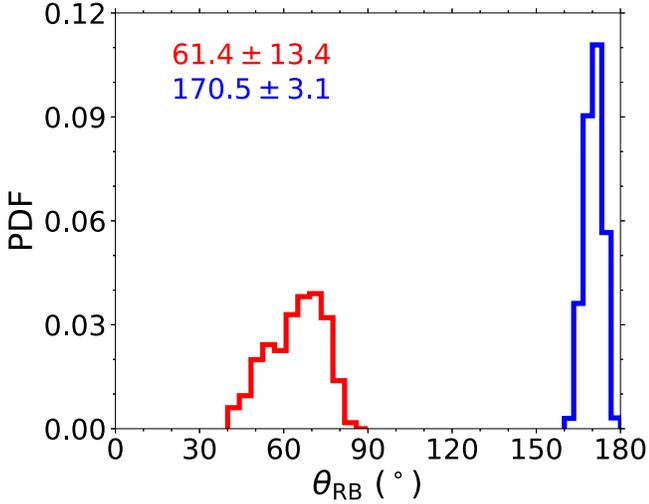

**Figure 3.** Probability distribution of the angle $\theta_{RB}$ for the selected switchback intervals (red) and the selected nonswitchback intervals (blue). The averages and standard deviations of $\theta_{RB}$ are shown in corresponding colors.

The left panel in Figure 5 demonstrates the probability distribution of $C_{vb}$. The distribution of $C_{vb}$ for the selected switchback intervals is shown in red with an average 0.65 and the distribution for the selected nonswitchback intervals is shown in blue with an average 0.53. The distribution of the switchbacks indicates an alignment between the magnetic fluctuations and the velocity fluctuations. The right panel in Figure 5, however, presents distinct distributions of $r_A$ between switchback intervals and nonswitchback intervals. $r_A$ is the ratio between kinetic and magnetic energy. The most probable value of $r_A$ for the switchbacks is 0.35–0.4 while for the nonswitchbacks is around 1.1–1.4. The distribution of $r_A$ for the switchbacks indicates that their nature may be magnetically dominated structures.

Figure 6(A1) illustrates the pixel average amplitude of $\delta v_A$ as a function of $C_{vb}$ and $\sigma_r$ for the switchback intervals. In each pixel, the color denotes the average amplitude of $\delta v_A$ for the intervals with the corresponding $C_{vb}$ and $\sigma_r$. It shows a large amplitude vertical stripe feature over the area $C_{vb} \in (0.8, 1.0)$ in the $C_{vb}$–$\sigma_r$ plane, and that the amplitudes of $\delta v_A$ increase with $C_{vb}$ and are nearly independent on $\sigma_r$. Figure 6 (A2) displays the pixel standard error $\sigma$ of $\delta v_A$ in the $C_{vb}$–$\sigma_r$ plane. The average amplitudes in each pixels are reliable since the standard errors are much smaller in the corresponding pixels. It should be noted that the standard error is zero in some pixels where the number is only one as shown by the pixel number distribution of switchback intervals (Figure 6(A5)). This problem is to be improved by more measurements in the future. Figures 6(A3) and (A4) illustrate the pixel average amplitude and pixel standard error of $\delta v$ as a function of $C_{vb}$ and $\sigma_r$ for the switchback intervals. A large amplitude vertical stripe feature also exists for $\delta v$ over the area $C_{vb} \in (0.8, 1.0)$. The vertical large amplitude stripe features appearing at high correlation $C_{vb} > 0.8$ between the magnetic field fluctuations and the velocity fluctuations are similar to the the stripe features of Magnetic-Velocity Alignment Structures (MVAS), which are identified at $|C_{vb}| > 0.85$ and $\sigma_r \in (-0.9, -0.2)$ by Wang et al. (2020). We should note here that most of the switchback events are located at $C_{vb} \in (0.5, 0.9)$ and $\sigma_r \in (-0.6, -0.2)$, broader than the results in $C_{vb}$ in Wang et al. (2020). The percentage of the switchbacks with $C_{vb} > 0.8$ is 24.4% of the total events.

Figures 6 (A1) and (A3) also tell us that the magnetic fluctuations are stronger than the kinetic fluctuations in the switchbacks. Figures 6 (B1) and (B3) show the pixel average amplitude of $\delta v_A$ and $\delta v$ for the nonswitchback intervals and Figures 6 (B2) and (B4) display the corresponding standard errors. The standard errors are small compared to the average

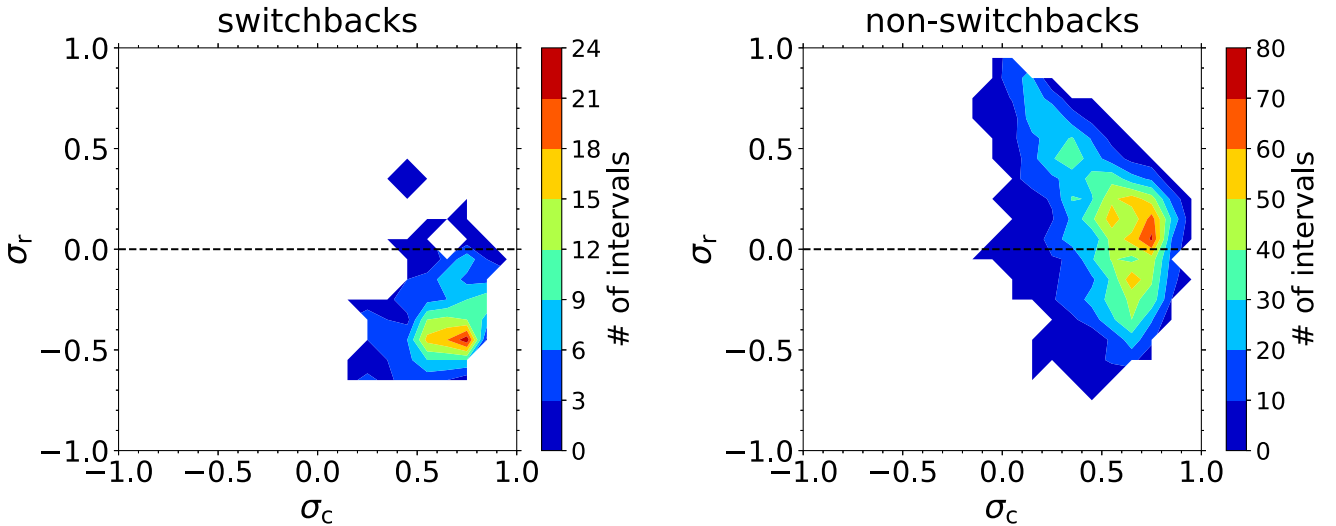

**Figure 4.** The number distribution in the ($\sigma_c$, $\sigma_r$) plane for the selected switchback intervals (left) and the selected nonswitchback intervals (right).

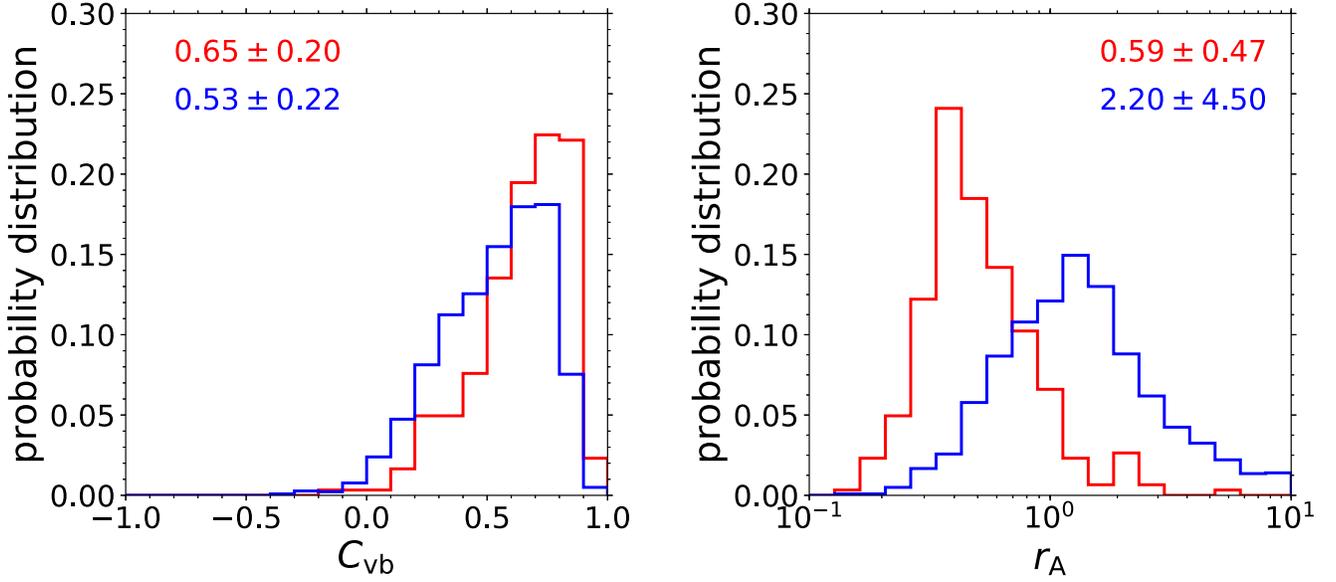

**Figure 5.** Left: probability distribution of $C_{vb}$ for the switchback intervals (red) and the nonswitchback intervals (blue). The averages and standard deviations of $C_{vb}$ are shown in corresponding colors. Right: probability distribution of Alfvén ratio $r_A$ in the same format as the left panel.

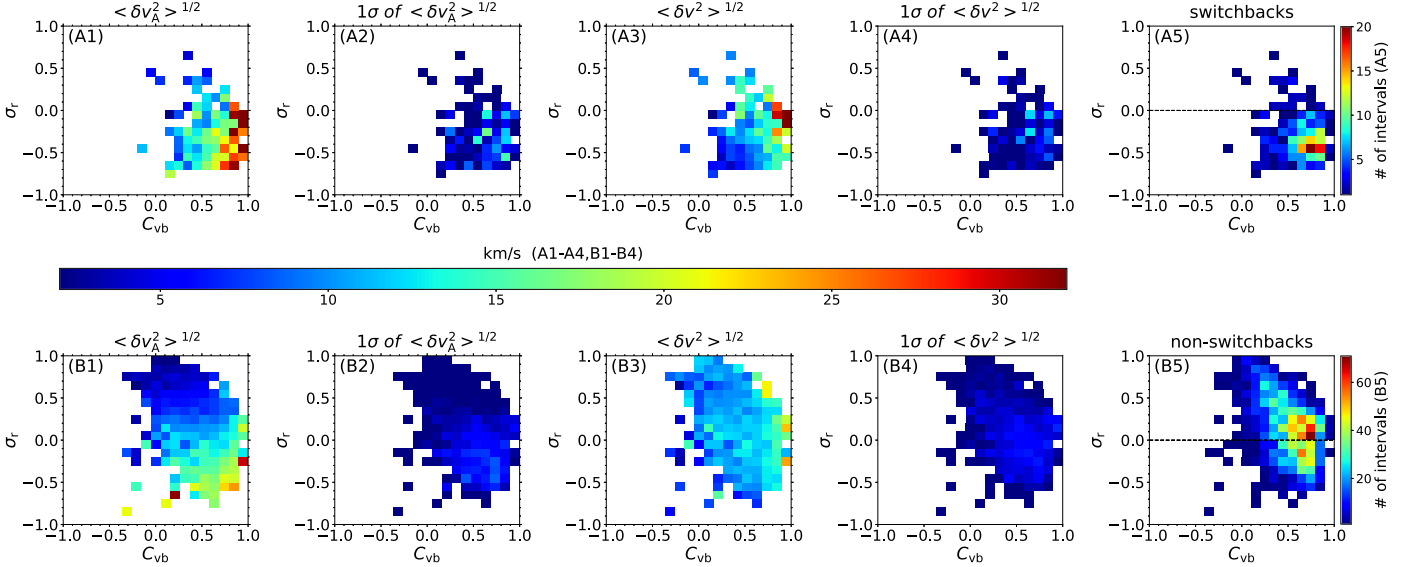

**Figure 6.** Top, from left to right (A1–A5): pixel average amplitude (A1) and pixel standard error (A2) of $\delta v_A$; pixel average amplitude (A3) and pixel standard error (A4) of $\delta v$; pixel number distribution (A5), respectively, in the $C_{vb}$–$\sigma_r$ plane for the selected switchback intervals. Bottom, from left to right (B1–B5): in the same format of A1–A5 for the selected nonswitchback intervals. A1–A4 and B1–B4 share the same colorbar plotted in the middle.

amplitudes in those pixels for the nonswitchback intervals, whose number distribution is shown in Figure 6(B5). It is clear that there are no vertical stripe features but rather oblique stripe features for $\delta v_A$. Comparing Figures 6(A1) and 6(B1), we can see that the magnetic fluctuations are larger in the switchbacks (Figure 6(A1)) than the corresponding fluctuations in the nonswitchbacks (Figure 6(B1)). It is also interesting to see that for those nonswitchbacks, the amplitude of kinetic fluctuations is nearly independent of $\sigma_r$ while the amplitude of magnetic fluctuations decrease as $\sigma_r$ increases. This suggests that the value of $\sigma_r$ could be determined by the amplitude of magnetic fluctuations but not kinetic fluctuations.

## 4. Conclusions and Discussion

We perform statistical analyses of turbulent state parameters for the fluctuations inside the switchbacks and inside the quiet solar wind with every interval duration of 100 s. In order to avoid the influences of ambiguous deflections and the switchback boundaries, we select the switchbacks and nonswitchbacks using the criteria $B_R > 0$ and $\theta_{RB} > 160°$, respectively. We find that the switchback intervals and nonswitchback intervals have similar $\sigma_c$ distributions but distinct $\sigma_r$ distributions, suggesting that switchbacks may have a different Alfvénic nature with respect to the Alfvén ratio from the nonswitchbacks. We show that the correlation coefficient $C_{vb}$ is as high as around 0.7 and the $r_A$ is around 0.35 in switchbacks, indicating that the magnetic field and velocity fluctuations in the switchbacks are closely aligned but magnetically dominated. We further find that the pixel average amplitudes of $\delta v_A$ and $\delta v$ in the $C_{vb}$–$\sigma_r$ plane in switchbacks present vertical stripe features.

In our study, we find that $\sigma_r$ in the switchbacks is negative, smaller than the ~zero value in the nonswitchbacks. This result confirms the finding of Bourouaine et al. (2020) that the residual energy of switchbacks is larger. However, we also find that the distributions of $\sigma_c$ show no difference between switchbacks and nonswitchbacks. This does not support the conclusion of Bourouaine et al. (2020) that switchbacks are more balanced than nonswitchbacks, which is based on one discontinuous interval analysis using the conditioned correlation functions technique.

Tu & Marsch (1994) pointed out that both the temperature variations and the flow speed variations may be considered as representing tiny flow structures or tubes, which are oriented along the mean magnetic field lines. Tu et al. (2016) presented the first two time series with high $C_{vb}$ values and low $r_A$ values, and suggested that these time series may represent convecting magnetic field structures in the solar wind. Wang et al. (2020) first suggested the concept of magnetic-velocity alignment structures. They are magnetically dominated structures with high correlation between magnetic field fluctuations and velocity fluctuations—MVAS, of which $C_{vb} \sim 0.8$–$0.9$ and $\sigma_r \sim (-0.9) - (-0.2)$. Our results show that large amplitude switchback turbulence with $C_{vb} > 0.8$ is likely MVAS. However, we should point out that a large portion of the cases with moderate or small amplitude and with low $C_{vb}$ are not in the range of MVAS. The nature of these fluctuations needs further study. A candidate may be the magnetic structures, reported by Wang et al. (2020). The magnetic-velocity alignment structures may have a flux-tube texture (Borovsky 2020), which are possible remnants of coronal jets (Sterling & Moore 2020). They could be small flux ropes injected by the interchange reconnection between coronal loop and open magnetic field in the corona (Fisk 2005; Drake et al. 2020; Fisk & Kasper 2020; Zank et al. 2020). Further investigation is required to uncover the nature and the formation of switchback turbulence.

We are thankful for helpful discussions with Dr. Jiansen He. We acknowledge the NASA Parker Solar Probe mission team and the SWEAP team led by J. C. Kasper, and the FIELDS team led by S. D. Bale, for the use of PSP data. The data used in this paper can be downloaded from spdf.gsfc.nasa.gov. This work is supported by the National Natural Science Foundation of China under contract Nos. 41974198, 41674171, 41874199, 41974171, and 41774157. H. W. is also supported by China Postdoctoral Science Foundation (2020M680204).

## ORCID iDs

Honghong Wu  https://orcid.org/0000-0003-0424-9228
Chuanyi Tu  https://orcid.org/0000-0002-9571-6911
Xin Wang  https://orcid.org/0000-0002-2444-1332
Liping Yang  https://orcid.org/0000-0003-4716-2958